\documentclass[aps,prb,twocolumn,showpacs,superscriptaddress]{revtex4}

\usepackage{graphicx} 
\begin{document}

\title
{\textbf{A first-principles study of GaAs nanowires}}
\author{S. Cahangirov}
\affiliation{UNAM-Institute for Materials Science and
Nanotechnology, Bilkent University, Ankara 06800, Turkey}
\author{S. Ciraci} \email{ciraci@fen.bilkent.edu.tr}
\affiliation{UNAM-Institute for Materials Science and
Nanotechnology, Bilkent University, Ankara 06800, Turkey}
\affiliation{Department of Physics, Bilkent University, Ankara
06800, Turkey}
\date{\today}

\begin{abstract}
In this paper we present a detailed analysis of the atomic and
electronic structure of GaAs nanowires using first-principles
pseudopotential calculations. We consider six different types of
nanowires with different diameters all grown along [111] direction
and reveal interesting trends between cohesive energy and nanowire
type with varying diameters. Generally, the average cohesive
energy of nanowires with wurtzite stacking is higher than those
with zinc blende stacking for small diameters. We found that most
of the bare nanowires considered here are semiconducting and
continue to be semiconducting upon the passivation of surface
dangling bonds with hydrogen atom. However, the surface states
associated with the surface atoms having two dangling bonds in
zinc blende stacking occur in the band gap and can decrease 
the band gap to change the nanowire from semiconducting to
metallic state. These nanowires become semiconducting upon
hydrogen passivation. Even if the band gap of some nanowires
decreases with increasing diameter and hence reveals the quantum
confinement effect, generally the band gap variation is rather
complex and depends on the type and geometry, diameter, type of
relaxation and also whether the dangling bonds of surface atoms
are saturated with hydrogen.

\end{abstract}


\maketitle

\section{Introduction}
In a search to discover semiconductor materials and metallic
interconnect for new generation miniaturized electronic devices,
nanostructures have been a focus of attention. Electronic devices,
such as transistors based on carbon nanotubes\cite{review},
attracted interest in nanowires. Rodlike Si nanowires have been
fabricated\cite{sinw1} with a diameter 1.3-7 nm.\cite{sinw2} It
has been shown that such Si nanowires can display metallic,
semiconducting and half-metallic properties depending on their
functionalization.\cite{sinw3} Being an alternative to silicon
based microelectronics GaAs is one of the most important materials
used in semiconductor physics. Due to the high electron mobility,
GaAs always carried a potential of being used in high speed
electronic devices. GaAs/AlGaAs
heterostructures\cite{esaki} have served as a media for the two
dimensional electron gas studies.

Similar to bulk crystals researcher have envisioned GaAs naowires
to be a potential alternative for Si nanowires. Recent advances in
fabrication technology made it possible to grow GaAs nanowires.
They are grown by metal catalysts in vapor-liquid-solid (VLS)
mechanism.\cite{harmand} Generally, GaAs nanowires are grown along
[111] direction in zincblende ($zb$) structure, whereas nanowires
with wurtzite ($wz$) structure with diameter as small as 10 nm are
also observed.\cite{mariager} Several models were developed to
predict the transition radius from wurtzite to zinc blende
structure.\cite{glas, dubrovskii} Actually, there is no sharp
transition but instead there are many different stacking
configurations with very similar energies and more sophisticated
models are needed to predict the ground state configuration.
Together with the crystal structure, surface facet structure is
also an important parameter affecting the structural and
electronic properties. Nanowires reported so far are either grown
in $zb$ structure with [$11\bar{2}$] or [$1\bar{1}0$] facet
orientation or in $wz$ structure with [$1\bar{1}00$] or [$11\bar{2}0$]
facet orientation.\cite{leitsmann, hiruma, ohlsson} One of the
unusual structures is so called A-wire, which has been grown in
defect free triangular shape on (111)A surface of
GaAs.\cite{wacaser} More complex structure of nanowires grown on
GaAs(111)B surface have been also reported.\cite{lisa}

There are many experimental and theoretical works on growth and
structure of GaAs nanowires, but not much work is done to predict
the electronic properties of these wires. Theoretical studies so
far have focused on the electronic properties of superlattices
composed of GaAs nanowires. For example, the electronic structure
of InAs/GaAs nanowire superlattices with radius R=10 nm was
examined using a semiempirical $sp^3d^5s^*$ tight-binding
model.\cite{niquet} Another atomistic tight-binding calculation
was carried out to reveal the electronic structure of freestanding
GaAs/Al$_{0.3}$Ga$_{0.7}$As nanowire superlattices oriented along
the [100] crystallographic direction.\cite{persson} Also a 
first-principles investigation has been performed on the
hexagon-shaped, [111]/[0001]-oriented III-V semiconductor
nanowires, which was concentrated on the surface effects on the
structure and stability of these nanowires.\cite{leitsmann}

The purpose of this paper is to provide a detailed analysis of
GaAs nanowires, which is necessary for further experimental and
theoretical studies. To this end we present a systematic,
first-principles investigation on structural and electronic
properties of GaAs nanowires grown along [111] direction. Six
different types of GaAs nanowires are distinguished depending on
the shape of their cross sections and the crystallographic
orientation of their side surfaces. Their optimized atomic
structure and cohesive energies are calculated revealing
interesting trends between atomic structure and cohesive energy.
Based on the calculations of electronic structure and isosurface
charge density of specific states we analyzed the character of
states at the band edges and variation of band gap with diameter.
The effects of hydrogen saturation of dangling bonds of surface
atoms on the atomic and electronic structure are examined. We
found that most of bare GaAs nanowires are semiconducting and
remain semiconducting even before the passivation of surface
dangling bonds. Only one type is metallic due to the states
localized at the surface.

\section{Methods}

We have performed first-principles plane wave
calculations\cite{payne, vasp} within Density Functional Theory
(DFT)\cite{kohn} using ultrasoft pseudopotentials.\cite{vasp,
vander} The pseudopotentials having three electrons for Ga (4$s^2$ 4$p^1$), 
five electrons for As (4$s^2$ 4$p^3$) and one electron for H (1$s^1$) were used. A plane-wave basis set with kinetic energy up to 250 eV was used. Cutoff energies used were at least 30 $\%$ higher than maximum values suggested.\cite{vasp} The exchange correlation potential is approximated by generalized gradient approximation (GGA) using PW91
functional.\cite{gga} For partial occupancies we use the
Methfessel-Paxton smearing method.\cite{methfessel} The adopted
smearing width is 0.1 eV for the atomic relaxation and 0.01 for
the accurate band structure analysis and density of state
calculations. All structures have been treated within a supercell
geometry using the periodic boundary conditions. Vacuum spacing was arranged so that the minimum distance between two atoms in adjacent unit cells were larger than 10 $\AA$, provided that atoms have negligible interaction at that far distances. We chose bare $wz3-96$ as a test structure and increased the vacuum spacings to 16 $\AA$. This resulted in an energy 
difference around 0.2 meV/atom. In the self-consistent 
potential and total energy calculations the
Brillouin zone (BZ) is sampled in the \textbf{k}-space within
Monkhorst-Pack scheme\cite{monk} by (1x1x9) mesh points for $wz$
and (1x1x7) mesh points $zb$ nanowires. Increasing the \textbf{k}-space sampling for bare $wz3-96$ structure from (1x1x9) to (1x1x15) resulted in a total energy difference around 1 meV. All atomic positions and lattice constant are optimized by using the conjugate gradient
method where total energy and atomic forces are minimized. The
criterion of convergence for energy is chosen as 10$^{-5}$ eV
between two ionic steps, and the maximum force allowed on each
atom is 0.05 eV/$\AA$. We have reduced the maximum force criterion 
down to 0.0025 eV/$\AA$ in our test structure, bare $wz3-96$. 
This had no considerable effect, since the change in energy and band gap was around 0.3 meV/atom and 0.006 eV, respectively. Clearly, the criterion for the maximum allowed force 0.05 eV/$\AA$ is appropriate for systems including large number of atoms.

\begin{figure}
\includegraphics[width=8cm]{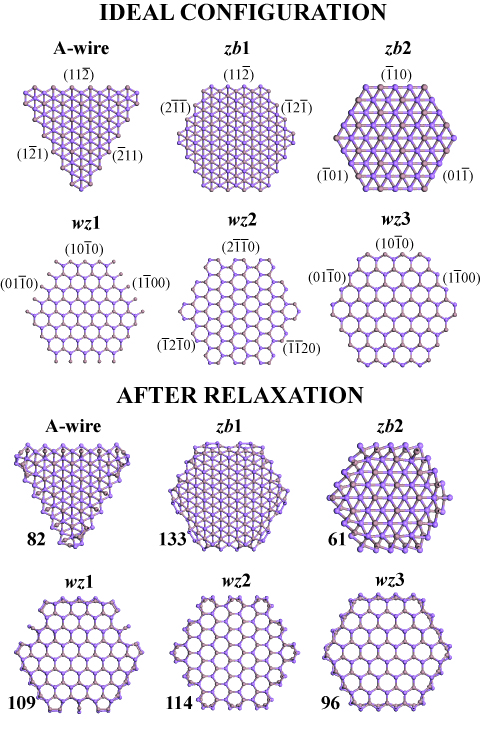}
\caption{(Color online) Ideal and relaxed atomic structures of
bare GaAs nanowires considered in this paper. Numerals given in
parenthesis indicate the crystallographic directions perpendicular
to the surfaces. Numerals given to the bottom left of the
structures stand for the number of atom pairs per unit cell $N$.
$wz$ and $zb$ stand for structures having wurtzite and zincblende
stackings.}\label{fig:fig1}
\end{figure}

\begin{figure}
\includegraphics[width=8cm]{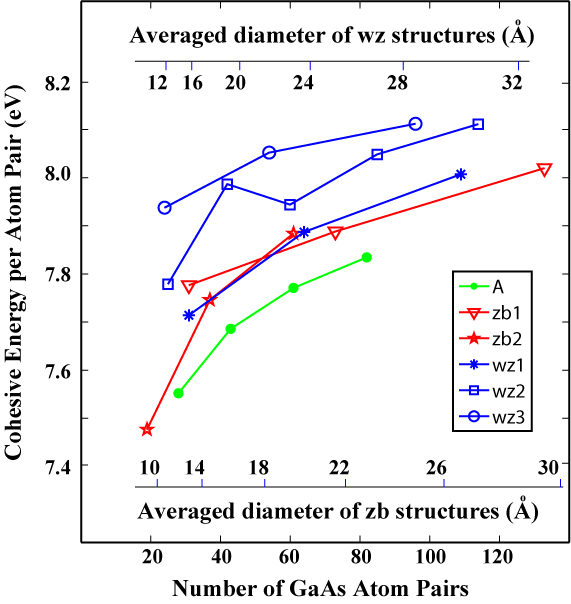}
\caption{(Color online) Cohesive energy per Ga-As atom pair versus
number of Ga-As atom pairs in the unit cell of different type of
relaxed nanowires. Horizontal axes presented inside the figure 
are derived by fitting the diameter versus number of atom data of $wz$ and $zb$ 
nanowires to quadratic polynomials. Since $wz$ and $zb$ structures 
have different number of atomic planes in the unit cell the fitting was done separately.}
\label{fig:fig2}
\end{figure}

\section{Structures and Cohesive Energies}

GaAs nanowires studied here are cut from ideal bulk structure
along [111] direction. Nanowires having wurtzite ($wz$) and zinc
blende ($zb$) stacking have four and six atomic layers in the unit
cell, respectively. Except A-wire, they have hexagonal cross
section. A-wires, by themselves have $zb$ stacking and display a
triangular cross-section with three ($11\bar{2}$) planar side
surfaces. Atomic structures of cross-section of all nanowires are
shown before and after relaxation (structure optimization) in 
Fig.~\ref{fig:fig1}. Here we consider the largest members of all
types of GaAs nanowires. Upon relaxation the surfaces of ideal wires
undergo a reconstruction, while inner parts preserve the bulk
configuration. In spite of the fact that the indices of their
planar side surface are the same, $wz1$ and $wz3$ are still
different. $wz3$ structure have six identical surfaces, whereas
$wz1$ structure have three surfaces same as $wz3$ and three
surfaces with hanging Ga-As atom pairs before relaxation.

At the surfaces of some ideal structures in Fig.~\ref{fig:fig1}
the atoms can have the coordination number smaller than four. 
Upon relaxation the coordination numbers may
undergo a change. Two adjacent surface atoms having low
coordination number can form new bonds, whereby these atoms
increase their coordination number and the nanowire, in turn,
lowers its energy (i.e it becomes more energetic). It turns out
that, the coordination number of surface atoms is crucial for the
value of the cohesive energy per atom pair. For example, $wz1$
nanowire with $N=109$ has three surfaces each having an atom pair
with coordination number two, while other three surfaces have
surface atoms with coordination number three. $wz2$ nanowires
having $N=$25, 60 and 85 atom pairs in the unit cell have surface
atoms with coordination number two only at the corners of the
hexagonal cross-section. On the other hand, $wz2$ nanowires having
$N=$42 and 114 atom pairs in the unit cell and all members of
$wz3$ nanowires have surface atoms with coordination number three.
In light of these arguments one expects $wz3$ to have larger
cohesive energy per atom pair than the rest of $wz2$ nanowires,
which, in turn, should be larger than that of $wz1$ nanowires.
These arguments are actually confirmed in Fig.~\ref{fig:fig2},
where we present the trends of cohesive energies per atom pair for
all nanowires considered here.

Ideal A-wires have surface atoms with coordination number of two
on each surface, but apart from that, it has a triangular
cross-section, which makes the surface to volume ratio even higher
compared to that of other types. Consequently, A-wires have the
lowest cohesive energy per atom pair as seen in Fig.~\ref{fig:fig2}.

\begin{table*}
\caption{Calculated cohesive energy per Ga-As atom pair $E_{c}$,
band gap $E_{g}$, lattice constant along the wire axis $c$ and
diameter $D$ values of relaxed nanowires are given. $D$ is defined
as the largest distance between two atoms in the same
cross-sectional plane. Here $N$ stands for the number of Ga-As
atom pairs in the unit cell. $N_S$ and $N_D$ stand for the number
of surface atoms and the total number of dangling bonds,
respectively. Surface atoms are defined as atoms making less than
four bonds, while the protruding bonds are defined as dangling
bonds.} \label{tab:table1}
\begin{tabular}{c|ccc|ccccc|ccc|cccc|ccc|ccc}
\hline\hline
Type&\multicolumn{3}{c|}{$wz1$}&\multicolumn{5}{c|}{$wz2$}&\multicolumn{3}{c|}{$wz3$}&\multicolumn{4}{c|}{A}&\multicolumn{3}{c|}{$zb1$}&\multicolumn{3}{c}{$zb2$}\\
\hline
$N$&31&64&109&25&42&60&85&144&24&54&96&28&43&61&82&31&73&133&19&37&61\\

$E_{c}$(eV)&7.71&7.89&8.01&7.78&7.99&7.94&8.05&8.11&7.94&8.05&8.11&7.55&7.69&7.77&7.83&7.78&7.89&8.02&7.48&7.75&7.88\\

$E_{g}$(eV)&1.06&1.02&0.92&0.81&0.96&0.95&0.92&0.88&1.45&1.10&0.90&0.81&0.77&0.85&0.70&1.07&0.14&0.58&M&M&M\\

$c($\AA$)$&6.64&6.63&6.63&6.58&6.59&6.61&6.61&6.63&6.60&6.61&6.63&9.94&9.89&9.88&9.87&9.97&9.95&9.85&9.97&9.87&9.84\\

$D($\AA$)$&15.9&23.6&31.6&14.5&18.1&23.1&28.4&32.0&12.5&20.5&28.6&12.6&16.3&19.8&24.5&12.7&20.9&29.0&9.2&14.5&18.3\\

$N_S$&24&36&48&24&36&36&48&60&24&36&48&29&36&45&54&30&48&66&24&36&48\\

$N_D$&30&36&54&27&36&43&51&60&24&36&48&38&42&54&66&30&54&66&30&42&54\\
\hline\hline
\end{tabular}
\end{table*}

Interestingly, $wz$ structures have relatively larger cohesive
energies than $zb$ structures, with $wz3$ structure being the
largest. The bulk $zb$ structure however, is energetically more
favorable than that of $wz$ structure by nearly 20 meV per Ga-As
pair, also confirmed by our calculations. As radii increase, the
cohesive energy per Ga-As atom pair values should converge to the
bulk value. It is energetically easier to form a $wz$ surface than
to form a $zb$ surface. In other words surface energy of $zb$
structure is larger. That is why, $wz$ structures become
energetically more favorable as the surface to volume ration
increases, namely as the radius of the nanowire decreases. So
there should be some point where cohesive energy of $wz$ and $zb$
structures cross each other. The radius at this point can be
interpreted as the critical radius for transition from $wz$ to
$zb$ structure or vice versa. Here it should be noted that,
energetically favorable does not mean that these structures will
start to grow in experiments. One should also include effects of
formation path, like nucleation growth. Actually, when we are
around the critical radius we can see some hybrid stacking
patterns like ABAC, which is defined as 4H
structure.\cite{dubrovskii}

\begin{figure}
\includegraphics[width=8cm]{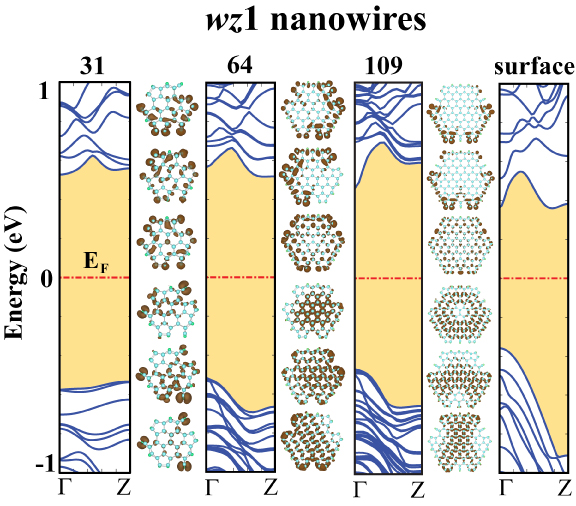}
\caption{Band structure and charge density isosurface plots of
GaAs nanowires having $wz1$ structure. Energy band gap between the
valance and conduction band is shaded. Numerals given on top of
the bands stand for the number of GaAs atom pairs $N$ in the unit
cell. Charge density isosurfaces of specific states at
$\Gamma$-point are shown on the right hand side of the bands they
belong to. Isosurface charge densities correspond to three valence
and three conduction band edge states, ordered in the same manner
as bands themselves are. Here we also give the band structure of
infinite slab of bulk wurtzite structure consisting 11 atomic
layers with the same planar $(10\bar{1}0)$ surfaces as $wz1$
nanowires does. Zero of energy is set at the Fermi level
$E_F$.}\label{fig:fig3}
\end{figure}

Table \ref{tab:table1} gives the calculated values for the
structure and cohesive energies of nanowires after the relaxation.
Cohesive energies per Ga-As atom pair increase with increasing
diameter, approaching the bulk values, but we don't see the
critical radius because it is expected to be an order of magnitude
larger than that of our nanowires.\cite{dubrovskii} Surprisingly,
the lattice constant decrease as the diameter of nanowires in $zb$
structure increase, while for $wz$ structures the reverse
situation occurs. The ratio of number of surface atoms to the
total number of atom pairs give a measure of surface to volume
ratio, which is decreasing with increasing diameter. Note that the number
of surface atoms having coordination number of two is
$(N_{D}-N_{S})$.

\begin{figure}
\includegraphics[width=8cm]{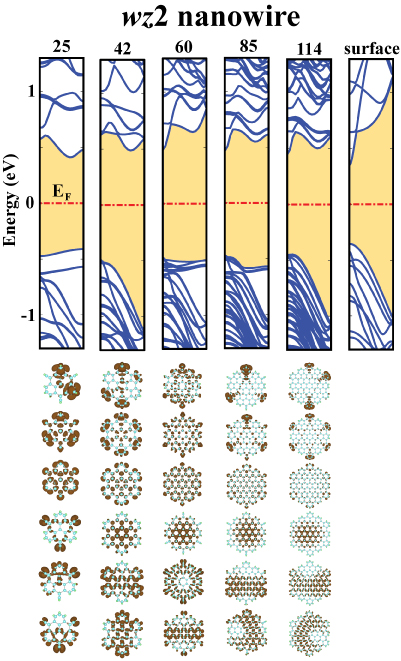}
\caption{Same as in Fig.3 but for $wz2$ structure. Isosurface
charge densities correspond to three valence and three conduction
band edge states, ordered in the same manner as bands themselves
are.}\label{fig:fig4}
\end{figure}

\section{Electronic Structure}

Most of the relaxed GaAs nanowires presented in
Fig.~\ref{fig:fig1} are semiconducting even without hydrogen
saturation. As we will see, in some cases these bare GaAs
nanowires don't even have the surface states at the band edges.
This situation is in contrast with Si
nanowires.\cite{sinw1,sinw2,sinw3,sinw4} Si nanowires as cut
from the bulk crystal and subsequently relaxed are found to be
metallic. Their metallicity occurs due to the partial filling of
the dangling bonds surface states. Upon passivation of the
dangling bonds with hydrogen atoms the surface states are
discarded from the band gap and eventually Si nanowire becomes
semiconductor. In what follows, we will examine the electronic
structure of bare GaAs nanowires and reveal the effects of geometry
and the passivation of dangling bonds with hydrogen.\cite{bandgap}

\begin{figure}
\includegraphics[width=8cm]{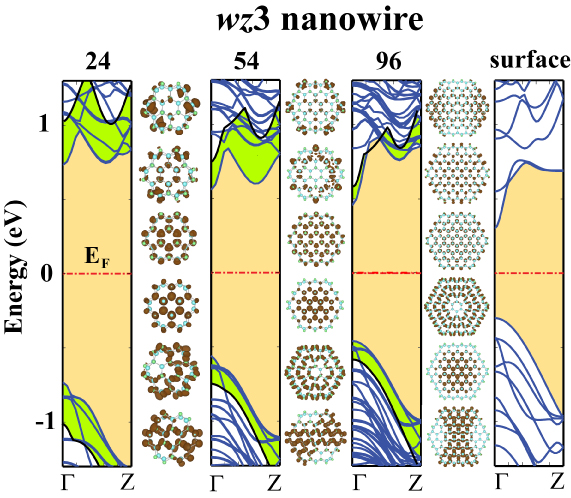}
\caption{Band structure and charge density isosurface plots of
GaAs nanowires having $wz3$ structure. Energy band gap between the
valance and conduction band of bare nanowire is (yellow)
light-shaded. Numerals given on top of the bands stand for the
number of GaAs atom pairs $N$ in the unit cell. Charge density
isosurfaces of specific states at $\Gamma$-point are shown on the
right side of the bands they belong to. Isosurface charge
densities correspond to three valence and three conduction band
edge states, ordered in the same manner as bands themselves are.
Here we also give the band structure of infinite slab of bulk
wurtzite structure consisting 11 atomic layers with the same
planar $(10\bar{1}0)$ surfaces as $wz3$ nanowires does. Zero of
energy is set at the Fermi level $E_F$. The widening of the band
gap upon the termination of dangling bonds by hydrogen is shown by
(green) dark-shaded regions delineated by black curves at the
valance and conduction band edges.} \label{fig:fig5}
\end{figure}

\begin{figure}
\includegraphics[width=8cm]{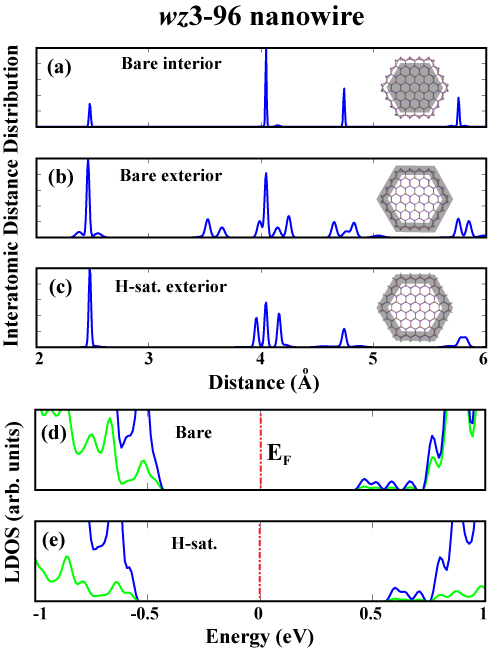}
\caption{Interatomic distance distribution of the core and shell
part of bare and hydrogen saturated $wz3-96$ structures. The ball
and stick model illustrates the structure of $wz3-96$ nanowire,
while the shaded regions defines the core and shell parts of the
nanowires. (a) Interatomic distance distribution of interior atoms of bare
nanowire. (b) Interatomic distance distribution of exterior atoms of bare
nanowires. (c) Interatomic distance distribution of hydrogen passivated
exterior atoms. (d) Local density of states (LDOS) on surface atoms 
(green/light) and on the remaining atoms (blue/dark) of bare nanowire. 
(e) Same as (d) after passivation of surface atoms with hydrogen.} \label{fig:fig6}
\end{figure}

Figure~\ref{fig:fig3} presents results of the band structure and
charge density analysis performed for $wz1$ structure. Ideal
structure of these wires have Ga-As atom pairs hanging on three
surfaces, while other three surfaces have the same profile as
$wz3$ type. Upon relaxation these hanging pairs tend to bend towards 
each other and lower energy by making unusual Ga-Ga and As-As
bonds. For example, $wz1$ structure having 64 atom pairs per unit
cell ($wz1-64$) have four hanging pairs before relaxation. After
relaxation first two and last two of them bend to each other and
form a stable structure. $wz1-31$ and $wz1-109$ nanowires have one
hanging pair after relaxation. As a result, all surface atoms of
$wz1-64$ structure have coordination number of three, while
$wz1-31$ and $wz1-109$ have six surface atoms with
coordination number of two (See Table~\ref{tab:table1}). 
Isosurface charge densities show that three states at
the top of the valence band edge of $wz1-64$ and $wz1-109$
structures have bulklike character, while conduction band edge
states concentrate at the surface. It is hard to define the nature 
of states in $wz1-31$ structure because it has low diameter. 
To calculate the band structure of the relevant surface 
we cut a slab from the bulk wurtzite structure parallel 
to $(01\bar{1}0)$ surfaces so that the resulting 
structure have 11 atomic layers. This slab has two
dimensional periodicity in the surface and a vacuum region between
adjacent surfaces, so that they don't interact. The band structure
of this slab plotted along \textbf{k}, parallel to [111]-direction, provide
us with information about the band structure of nanowires if they
were grown thick enough to have a reasonable bulk region in order
to reduce the corner effects. One expects the band gap of the
infinite slab structure to be lower than that of the nanowires.
Comparison of the bands of $wz1-109$ with those of $(01\bar{1}0)$
surface confirm their similarity and expected size effect.

Figure~\ref{fig:fig4} shows the band structure and charge density
analysis for $wz2$ structures. Here all surface atoms of $wz2-42$
and $wz2-114$ structures have coordination number of three, while
the rest of considered $wz2$ structures have surface atoms with
coordination number of two at the corners. Note that the valence
band edge of $wz2-42$ and $wz2-114$ structures mimic that of
infinite slab structure, while other structures fail to do that.
We see that the band gap of $wz2-42$ is larger than that of
$wz2-114$ structure in agreement with the quantum confinement
effect. On the other hand, however, $wz2-25$ structure have the
lowest band gap. This is caused by large surface to volume
ratio, which make surface effects pronounced. Isosurface charge
densities show that the valence band edge have bulklike character,
while conduction band edge states concentrate on the surface.

Band structure and charge density analysis of $wz3$ structure is
illustrated in Fig.~\ref{fig:fig5}. This structure has the largest
cohesive energy per atom value compared to other types of
nanowires in Fig.~\ref{fig:fig1}. In this nanowire all surface
atoms make three bonds and surface states derived therefrom do not
occur in the band gap. Consequently, the band structure of the
related infinite slab is very similar to that of $wz3-96$
nanowire. Charge density of the states at both band edges are
spread throughout the nanowire cross-section showing the bulk
character. Since band gap is not diminished by surface states
bands, one can clearly see the quantum confinement effect in these
nanowires.

\begin{figure}
\includegraphics[width=8cm]{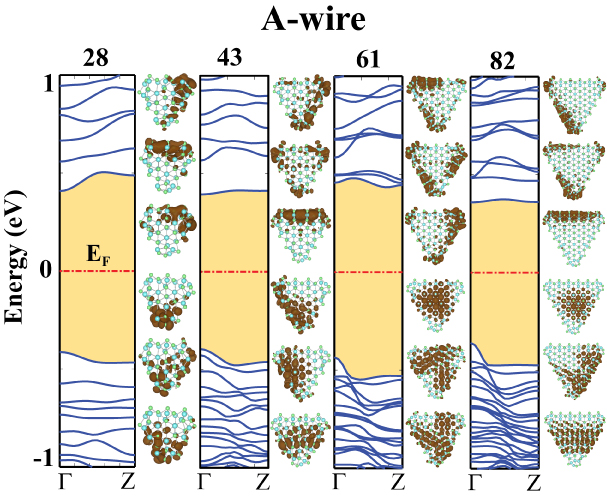}
\caption{Band structure and charge density isosurface plots for
A-wires.}\label{fig:fig7}
\end{figure}

Passivation of dangling bonds of semiconducting nanowires by
hydrogen atoms, generally, results in significant changes in the 
electronic structure. These changes depend on whether the passivation 
is done before or after the relaxation of bare nanowires. We find 
the latter case more suitable for the simulation of the experimental procedure.\cite{sinw3} Figure~\ref{fig:fig5} includes information about the effect
of hydrogen passivation of all surface dangling bonds on the band
structure of $wz3$ nanowires. In contrast to Si nanowires (where
surface states of bare structure were carried out from forbidden
region to the band continua upon passivation with
hydrogen)\cite{sinw3,ref1,ref2}, the band edge states of $wz3$ GaAs
nanowires remain in their place after the hydrogen saturation.
This conclusion is corroborated by the analysis of isosurface
charge density of states at the center of BZ located
at both edges of band gap. We found that the character and charge
distribution of these states do not undergo a change after
passivation of surface dangling bonds with hydrogen atoms. Also
the similarity in the profile of band edges before and after
hydrogen saturation is found to be striking. Hydrogen atoms mostly affect the surface states, which, in $wz3$ structure, are found in the valence band continua. That is why, the effect of hydrogen saturation is not reflected on the band edges in the way it was in silicon nanowires having surface states at the band edges. Here the increase of the band gap occur not because the edge states are cleared out, but because the atomic structure of the nanowires become more bulklike. This effect is illustrated in Fig.~\ref{fig:fig6}, where we analyzed the interatomic distance distribution of $wz3-96$ structure before and after the hydrogen saturation. Plots given here are done by making a histogram of interatomic distances and then smearing it out. The first plot indicates the interatomic distance distribution of the core region of bare nanowire. We get the same result also in the core region of hydrogen saturated
structure and the peaks match the first, second, third and fourth
nearest neighbor distances of bulk GaAs in $wz$ structure. The
crucial difference between bare and hydrogen saturated structures
is reflected in the shell part of the nanowires. One can clearly
see that hydrogen saturated structure have more bulklike
character.

Furthermore we performed atom projected density of states analysis
by calculating the localized density of states on the surface
atoms, as well as on the core atoms before and after passivation
with hydrogen atoms. As seen in Fig.~\ref{fig:fig6} (d) and (e),
surface as well as core atoms have comparable contributions to the
state distribution at both edges of band gap. These results further
corroborate the fact that GaAs nanowires like $wz3-96$ having
surface atoms with coordination number of three do not have
dangling bond surface states in the band gap. For nanowires with
large $D$ the passivation with hydrogen have negligible effects on
the band gap. Note that the increase in the band gap with decreasing
diameter holds also for hydrogen saturated structures.

\begin{figure}
\includegraphics[width=8cm]{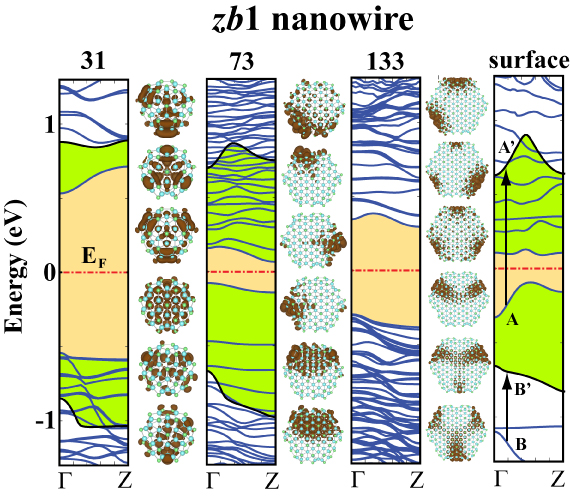}
\caption{(Color online)Band structure and charge density
isosurface plots of GaAs nanowires having $zb1$ structure. Energy
band gap between the valence and conduction band of bare nanowire
is (yellow) light-shaded. Numerals given on top of the bands stand
for the number of GaAs atom pairs $N$ in the unit cell. Charge
density isosurfaces of specific states at $\Gamma$-point are shown
on the right hand side of the bands they belong to. Isosurface
charge densities correspond to three valence and three conduction
band edge states, ordered in the same manner as bands themselves
are. Here we also give the band structure of infinite slab of bulk
wurtzite structure with the same planar $(11\bar{2})$ surfaces as
$zb1$ nanowires does. Zero of energy is set at the Fermi level
$E_F$. The widening of the band gap upon the termination of
dangling bonds by hydrogen is shown by (green) dark-shaded regions
for $N=31$, 73 and surface.}\label{fig:fig8}
\end{figure}

In Fig.~\ref{fig:fig7} we present the band structures and
isosurface charge densities of selected states at $\Gamma$-point
of A-wires. Note that, we cannot have an infinite slab
corresponding to the large A-wires because these nanowires have
triangular cross-section. All relaxed A-wires have surface atoms
with coordination number of two. The band gap have no obvious
trend with varying diameter and is determined by surface states
especially falling in the conduction band edge. For small $D$ the
edges of band gap are determined by surface states with low
dispersion. As the diameter increases the valence band edge starts
to show bulk character, while states at the conduction band edge
remain to display surface character. Even for $A-82$ nanowire the
flat band states at the edge of the conduction band are located on
one of the planar side surface.

The effect of surface states on the band gap is dramatic for the
bare GaAs nanowires having $zb$ stacking. The values of band gap
as large as 1.5 eV occurring in $wz$-type nanowires reduce to the
values as small as $\sim$0.2 eV in $zb1$-type nanowires. The
calculated band gap can even be closed in $zb2$-type nanowires. It
appears that many of the dangling bond surface states, in
particular those associated with the surface atoms having
coordination number of two, are placed in the band gap. Therefore
one expect dramatic changes in the electronic properties after the
passivation of surface dangling bonds with hydrogen atoms. The
electronic structures of bare $zb1$-type nanowires are illustrated
in Fig.~\ref{fig:fig8}. Here all surface atoms of $zb1-31$ and
$zb1-113$ nanowires have coordination number of three after
relaxation, but certain surface atoms of $zb1-73$ structure cannot
find a pair to form a bond and stay with coordination number of
two. As a result the band gap of $zb1-73$ does not have a value
between that of $zb1-31$ and $zb1-133$, as one would expect. To
get more information about the nature of the states at both edges
of the band gap we have saturated all dangling bonds of $zb1-31$
and $zb1-73$ structures by hydrogen. After the hydrogen saturation
the band edge profile does not stay the same, as it was in the
case of $wz3$ structures. This means that the band gap of bare
nanowire has increased upon hydrogen passivation of dangling
bonds, since the band gap is cleared from the surface states.
Apparently, the band edge states of the bare $zb1$ nanowire
originate from the surface, while in $wz3$ structure they have
bulklike behavior. The different behavior of triply coordinated surface atoms in $zb1$ and $wz3$ structure is attributed to their structural orientation. Surface atoms of $zb1$ structure make atomic rings perpendicular to the wire axis. Dangling bonds of these atomic rings form several minibands (with low dispersion along $\Gamma$Z) in the band gap, with charge density localized at the surface. Surface atoms of $wz3$ structure, however, form atomic chains along the wire axis. The dangling bonds of these atomic chains form surface states with high dispersion, which fall in the band continua. The effect of hydrogen saturation is investigated in the case of a slab in Fig.~\ref{fig:fig8}. Here the first and third valence band edge states of bare slab structure are labeled as A and B, respectively. These dispersive states have bulklike character. 
Upon the passivation band A rises to A' forming the conduction band edge, and band B rises to B' forming the valence band edge of the passivated structure. Surface states having low dispersion are lowered to valence band continua.

\begin{figure}
\includegraphics[width=8cm]{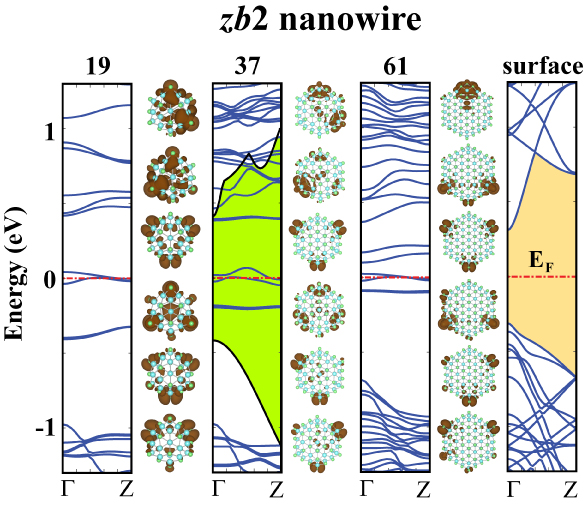}
\caption{Same as in Fig.~\ref{fig:fig8} but for $zb2$ structure.
The shaded region in the band structure of $zb2-37$ corresponds to
the band gap opened after the passivation of dangling bonds with
hydrogen. $zb2-19$ and $zb2-61$ nanowires are not passivated with hydrogen}\label{fig:fig9}
\end{figure}

The effects of the dangling bond surface states are even more
dramatic in $zb2$-type bare GaAs nanowires in Fig.~\ref{fig:fig9},
where the band gaps are closed and the nanowires become metallic.
Therefore, $zb2$-type is the only type we have considered to have
metallic character. It is interesting that the corresponding
infinite slab structure have a finite band gap. $zb2$ structures
would have a band gap which is larger than that of the slab
structure if they would had no partly filled corner states
crossing the Fermi level. Isosurface charge density plots support
the idea of band edge states being originated from the dangling
bonds at the corners of hexagonal cross-section. We further tested
these arguments by passivating the dangling bonds of $zb2-37$ with
hydrogen. Upon hydrogen passivation all flat surface state bands
in the range of energy from -0.5 eV to 0.5 eV disappeared and a band
gap of 0.9 eV opened. The resulting band profiles near the edges
of conduction and valence bands are similar to those of the slab
$(\bar{1}10)$ slab surface given at the right hand side of
Fig.~\ref{fig:fig9}.

\section{Discussions and Conclusions}

We have performed first-principles DFT calculations to reveal the
atomic and electronic structure of six different types of bare and 
hydrogen saturated GaAs nanowires. Nanowires considered have a diameter less than 3 nm,
and at this sizes $wz3$-type have the highest cohesive energy per
atom pair, while A-wires have the lowest. In general, $wz$
nanowires have higher cohesive energy than $zb$ nanowires, but the
difference in cohesive energies decrease with increasing diameter.

We found that all bare GaAs nanowires are semiconducting, except $zb2$-type. In the latter structure, dangling bond states of atoms having coordination number of two cross the Fermi level and hence the structure becomes metallic. For bare GaAs nanowires in $wz$ structure with surface atoms all having coordination number of three, the dangling bond states associated with these surface atoms do not appear in the band gap, but rather in the band continua. Therefore, quantum (size) confinement effect is apparent in their band gap variation with radius. For these $wz$-type nanowires the band gap may increase upon the passivation of dangling bonds with hydrogen, since bonding of surface atoms become more bulklike. 

Surface states of GaAs nanowires in $zb$ structure fall in the band gap, even if all surface atoms have coordination number of three. In $wz$ structure surface atoms form atomic chains along the wire axis, while in $zb$ structure they form non-interacting atomic rings perpendicular to the wire axis. Hydrogen saturation dramatically increases the band gap of $zb$ structures by clearing the surface states from the forbidden region. Generally, increasing diameter results in more bulk like valence band edge, however conduction band edge behave more surface like.

For reasons explained in detail, the band gap variation of GaAs nanowires is rather complex and
depends on their type and geometry, diameter, relaxation and also
whether the dangling bonds of surface atoms are passivated with
hydrogen. We believe that present results are valuable for further
research on GaAs and other III-V compound nanowires dealing with
their doping, forming heterostructure and multiple quantum well
structure and their fuctionalization to get new electronic and
magnetic properties.\cite{gudiksen,bjork}

\begin{acknowledgments}
Part of the computations have been carried out by using UYBHM at
Istanbul Technical University through a grant (2-024-2007).
\end{acknowledgments}


\begin{thebibliography}{99}
\bibitem{review} See for a recent review of the subject: S. Ciraci, S. dag, T. Yildirim. T. Senger and O Gulseren, J. Phys. Condens. Matter,\textbf{16}, R901 (2004)
\bibitem{sinw1} J. Hu, T.W. Odom and C.M. Lieber, Acc. Chem. Res.
\textbf{32}, 435 (1999)
\bibitem{sinw2} D.D.D. Ma, C.S. Lee, F.C.K. Au, S.Y. Tong and S. Lee, Science \textbf{299} 1874 (2003).
\bibitem{sinw3} E. Durgun, D. Cakir, N. Akman and S. Ciraci, Phys. Rev. Lett. \textbf{99}, 256806 (2007); E. Durgun, N. Akman and S. Ciraci, Phys. Rev. B \textbf{78}, 195116, (2008).
\bibitem{esaki}  L. Esaki and L.L. Chang, Phys. Rev. Lett. \textbf{33}, 495 (1974).
\bibitem{harmand}  J.C. Harmand, G. Patriarche, N. P�r�-Laperne, M-N. M�rat-Combes, L. Travers, and F. Glas, Applied Physics Letters, \textbf{87}, 203101, (2005).
\bibitem{mariager}  S. O. Mariager, C. B. S�rensen, M. Aagesen, J. Nyg�rd, and R.Feidenhans�l, Applied Physics Letters, \textbf{91}, 083106,(2007).
\bibitem{glas}  F. Glas, J.C.  Harmand, and G. Patriarche, Phys. Rev. Lett. \textbf{99}, 146101, (2007).
\bibitem{dubrovskii}  V.G. Dubrovskii and N.B. Sibirev, Phys. Rev. B \textbf{77}, 035414, (2008).
\bibitem{leitsmann}  R. Leitsmann and F. Bechstedt, J. Appl. Phys. \textbf{102}, 063528, (2007).
\bibitem{hiruma}  K. Hiruma, M. Yazawa, K. Haraguchi, and K. Ogawa, J. Appl. Phys. \textbf{74}, 3162 (1993).
\bibitem{ohlsson}  B.J. Ohlsson, M.T. Bj�rk, M.H. Magnusson, K. Deppert, and L.Samuelson, Appl. Phys. Lett. \textbf{79}, 3335 (2001).
\bibitem{wacaser}  Brent A. Wacaser, Knut Deppert, Lisa S. Karlsson, Lars Samuelson, Werner Seifert, Journal of Crystal Growth, \textbf{287}, 504 (2006).
\bibitem{lisa}  L.S. Karlsson, K.A. Dick, J.B. Wagner, J.O. Malm, K. Deppert, L. Samuelson and L.R. Wallenberg, Nanotechnology \textbf{18}, 485717 (2007).
\bibitem{niquet}  Y.M. Niquet, Phys. Rev. B \textbf{74}, 155304, (2006).
\bibitem{persson}  M.P. Persson and H.Q. Xu, Phys. Rev. B \textbf{73}, 035328, (2006).
\bibitem{payne}  M. C. Payne, M. P. Teter, D. C. Allen, T. A. Arias, and J. D. Joannopoulos, Rev. Mod. Phys. \textbf{64}, 1045 (1992).
\bibitem{vasp}  Numerical computations have been carried out by using VASP software: G. Kresse, J. Hafner, Phys Rev. B \textbf{47}, R558 (1993); G. Kresse, J. Furthmuller, Phys Rev. B \textbf{54}, 11169 (1996).
\bibitem{kohn}  W. Kohn and L. J. Sham, Phys. Rev. 140, A1133 (1965);  P. Hohenberg and W. Kohn, Phys. Rev. B 76, 6062 (1964).
\bibitem{vander}  D. Vanderbilt, Phys. Rev. B \textbf{41}, R7892 (1990).
\bibitem{gga}  J. P. Perdew, J. A. Chevary, S. H. Vosko, K. A. Jackson, M. R. Pederson, D. J. Singh, and C. Fiolhais, Phys. Rev. B \textbf{46}, 6671 (1992).
\bibitem{methfessel}  M. Methfessel and A. T. Paxton, Phys. Rev. B \textbf{40}, 3616 {1989}.
\bibitem{monk}  H.J. Monkhorst and J.D. Pack, Phys. Rev. B \textbf{13}, 5188, (1976).
\bibitem{sinw4}   P. Sen, O. Gulseren, T. Yildirim, I.P. Batra and S. Ciraci, Phys. Rev. B \textbf{65}, 235433 (2002).
\bibitem{sinw4}  R. Rurali and N. Lorente, Phys. Rev. Lett. \textbf{94}, 026805 (2005).
\bibitem{bandgap} Caution has to be taken in the band gap values
calculated in this study, since DFT underestimates their values.
For example, present calculations predict the band gap of bulk
GaAs only 0.27 eV as compared to the experimental value of 1.4 eV.
\bibitem{ref1} P. Leu, B. Shan, and K. Cho, Phys. Rev. B \textbf{73}, 195320 (2006).
\bibitem{ref2} M. Nolan, S. O'Callaghan, G. Fagas, et al. Nano Letters, \textbf{7}, 34-38 (2007).
\bibitem{gudiksen}
M. S. Gudiksen, L. J. Lauhon, J. Wang, D. C. Smith and C. M. Lieber, Nature {\bf 415}, 617 (2002).
\bibitem{bjork}
M. T. Bjork, B. J. Ohlsson, T. Sass, A. I. Persson, C. Thelander, M. H. Magnusson, K. Deppert, L. R. Wallenberg, and L. Samuelson, Nano Lett.{\bf 2}, 87 (2002).
\end{thebibliography}
\end{document}